\documentclass[notitlepage,nofootinbib,preprintnumbers,amssymb,superscriptaddress]{revtex4-1}

\usepackage{amsfonts,amssymb,amsmath,graphicx,color,bm}
\definecolor{ultramarine}{rgb}{0.07, 0.04, 0.56}
\definecolor{cadmiumgreen}{rgb}{0.0, 0.42, 0.24}
\definecolor{indigo(dye)}{rgb}{0.0, 0.25, 0.42}
\usepackage[linktocpage=true]{hyperref}
\hypersetup{
colorlinks=true,
citecolor=ultramarine,
linkcolor=cadmiumgreen,
urlcolor=indigo(dye),
}

\newcommand{\fr}[2]{\frac{#1}{#2}}
\newcommand{\pa}{\partial}
\newcommand{\ti}{\tilde}
\newcommand{\na}{\nabla}
\newcommand{\da}{\dagger}
\newcommand{\bra}[1]{\left( #1 \right)}  
\newcommand{\brb}[1]{\left[ #1 \right]}  
  
\newcommand{\be}{\begin{equation}}  
\newcommand{\ee}{\end{equation}}
\newcommand{\bem}{\begin{bmatrix}}
\newcommand{\eem}{\end{bmatrix}}
\newcommand{\Mpl}{M_{\rm Pl}}
\newcommand{\al}{\alpha}
\newcommand{\ga}{\gamma}
\newcommand{\ep}{\epsilon}

\newcommand{\la}{\lambda}
\newcommand{\si}{\sigma}

\newcommand{\mE}{\mathcal{E}}

\begin{document}

\preprint{RESCEU-1/17, RUP-17-2}

\title{General invertible transformation and physical degrees of freedom}

\author{Kazufumi Takahashi}
% \email{ktakahashi(at)resceu.s.u-tokyo.ac.jp}
\affiliation{Research Center for the Early Universe (RESCEU), Graduate School of Science, The University of Tokyo, Tokyo 113-0033, Japan}
\affiliation{Department of Physics, Graduate School of Science, The University of Tokyo, Tokyo 113-0033, Japan}
\author{Hayato Motohashi}
% \email{motohashi(at)ific.uv.es}
\affiliation{Research Center for the Early Universe (RESCEU), Graduate School of Science, The University of Tokyo, Tokyo 113-0033, Japan}
\affiliation{Instituto de F\'isica Corpuscular (IFIC), Universidad de Valencia-CSIC, E-46980, Valencia, Spain}
\author{Teruaki Suyama}
% \email{suyama(at)resceu.s.u-tokyo.ac.jp}
\affiliation{Research Center for the Early Universe (RESCEU), Graduate School of Science, The University of Tokyo, Tokyo 113-0033, Japan}
\author{Tsutomu Kobayashi}
% \email{tsutomu(at)rikkyo.ac.jp}
\affiliation{Department of Physics, Rikkyo University, Toshima, Tokyo 171-8501, Japan}

\begin{abstract}
An invertible field transformation is such that the old field variables correspond one-to-one to the new variables.
As such, one may think that two systems that are related by an invertible transformation are physically equivalent.
However, if the transformation depends on field derivatives, the equivalence between the two systems is nontrivial due to the appearance of higher derivative terms in the equations of motion.
To address this problem, we prove the following theorem on the relation between an invertible transformation and Euler-Lagrange equations: If the field transformation is invertible, then any solution of the original set of Euler-Lagrange equations is mapped to a solution of the new set of Euler-Lagrange equations, and vice versa.
We also present applications of the theorem to scalar-tensor theories.
\end{abstract}

\maketitle

%%%%%%%%%%%%%%%%%%%%%%%%%%%%%%%%%%%%%%%%%%%%%%%%%%%%%%%%%%%%%%%%%%%%%%%%%%%%%%%%%%%%
%%%%%%%%%%%%%%%%%%%%%%%%%%%%%%%%%%%%%%%%%%%%%%%%%%%%%%%%%%%%%%%%%%%%%%%%%%%%%%%%%%%%
%	Introduction
%%%%%%%%%%%%%%%%%%%%%%%%%%%%%%%%%%%%%%%%%%%%%%%%%%%%%%%%%%%%%%%%%%%%%%%%%%%%%%%%%%%%
%%%%%%%%%%%%%%%%%%%%%%%%%%%%%%%%%%%%%%%%%%%%%%%%%%%%%%%%%%%%%%%%%%%%%%%%%%%%%%%%%%%%
\section{Introduction}\label{sec:introduction}

It had been believed that the Horndeski theory~\cite{Horndeski:1974wa}, also known as the generalized Galileon theory~\cite{Deffayet:2011gz,Kobayashi:2011nu}, is the most general healthy (single-field) scalar-tensor theory.
This was because it forms the broadest class that yields second-order Euler-Lagrange~(EL) equations both for the metric and the scalar field, and thus trivially evade so-called Ostrogradsky ghosts associated with higher-order equations of motion~\cite{Woodard:2015zca}.
However, the myth was destroyed:
The second-order nature of the EL equations is just a sufficient condition and not a necessary condition for the absence of Ostrogradsky ghosts~\cite{Zumalacarregui:2013pma}.
Gleyzes, Langlois, Piazza, and Vernizzi~(GLPV) then constructed a healthy theory beyond Horndeski~\cite{Gleyzes:2014dya}.
The key for healthy theories beyond Horndeski is that the EL equations are {\it a priori} of higher order, but can be rearranged into a second-order system~\cite{Deffayet:2015qwa}, which is realized in the presence of appropriate degeneracy conditions or a sufficient number of constraints~\cite{Motohashi:2014opa,Langlois:2015cwa,Motohashi:2016ftl}.
Thus far many efforts have been made to construct healthy theories beyond Horndeski, which include quadratic/cubic degenerate higher-order scalar-tensor~(DHOST) theories~\cite{Langlois:2015cwa,BenAchour:2016fzp} and extended Galileons~\cite{Gao:2014soa,Fujita:2015ymn}.

Along this line, the transformation properties of these theories under the disformal transformation have been investigated in Refs.~\cite{Bettoni:2013diz,Zumalacarregui:2013pma,Gleyzes:2014dya,Gleyzes:2014qga,Crisostomi:2016czh,Achour:2016rkg,BenAchour:2016fzp,Fujita:2015ymn}.
A disformal transformation is a kind of frame transformation that generalizes conformal transformation.
The law of transformation is defined by
	\be
	\ti{g}_{\mu\nu}=A(\phi,X)g_{\mu\nu}+B(\phi,X)\na_\mu\phi\na_\nu\phi,~~~\ti{\phi}=\phi, \label{disformal}
	\ee
where $X\equiv -(\na_\mu\phi)^2/2$, and the functions $A,B$ are chosen so that the transformation does not change the metric signature and is consistent with the existence of the inverse matrix of $\ti{g}_{\mu\nu}$~\cite{Bekenstein:1992pj,Bettoni:2013diz}.
It was shown in Ref.~\cite{Zumalacarregui:2013pma} that there exists the inverse transformation of Eq.~\eqref{disformal} if $A,B$ satisfy some additional condition (see \S \ref{ssec:disformal} for detail).
One naturally expects that the number of physical degrees of freedom~(DOFs) is not changed by such an invertible transformation because there is a one-to-one correspondence between the old and new sets of variables.
On the other hand, since the disformal transformation contains derivatives of the scalar field, the EL equations derived from the transformed action contain higher-order derivatives, and thus the equivalence between the two frames is not clear.
There are some works that addressed this issue:
In the special case where the original action is of the Einstein-Hilbert form, the authors of Refs.~\cite{Zumalacarregui:2013pma,Deruelle:2014zza} showed that the EL equations in the new frame containing higher-order derivatives can be recomposed to yield second-order equations.
The disformal invariance of cosmological perturbations and their number of DOFs are investigated in Refs.~\cite{Minamitsuji:2014waa,Tsujikawa:2014uza,Watanabe:2015uqa,Motohashi:2015pra}, and it was clarified in Ref.~\cite{Domenech:2015hka} that the disformal transformation in a cosmological setup amounts to a rescaling of time coordinate, and thus leaves physical observables unchanged.
The authors of Ref.~\cite{Arroja:2015wpa} proved the equivalence between two sets of EL equations for disformally related frames for an arbitrary scalar-tensor theory.
The equivalence between the old and new frames has also been confirmed by Hamiltonian analysis in the unitary gauge~$\phi=t$~\cite{Domenech:2015tca}, though the similar analysis without gauge fixing remains unaddressed.
They also showed that the Hamiltonian structure is unchanged under a broad class of invertible field transformations.
However, there exist infinite different types of invertible transformations that are not covered by their analysis.
Also, the equivalence between EL equations for two frames related through general transformations has not been clarified.
These facts motivate us to explore the nature of generic invertible transformations that depend on fields and their derivatives.

Besides the above, there is another motivation to address the issue on invertible transformations:
It is related to {\it noninvertible} transformations, i.e., a class of transformations that are {\it not} invertible.
The so-called mimetic gravity model~\cite{Chamseddine:2013kea} is known as an example of a theory resulting from such a noninvertible transformation.
This theory is obtained by performing a particular noninvertible disformal transformation on the Einstein-Hilbert action in general relativity, and was shown to have three DOFs \cite{Chaichian:2014qba}.
Hence, in this case the noninvertible transformation increases the number of DOFs by one.
In general, a noninvertible transformation could map a theory to one with a different number of DOFs as opposed to the case of invertible transformations.
It is intriguing to investigate how the number of DOFs of a given theory changes by a generic noninvertible transformation.
Although this problem is beyond the scope of the present paper, the methodology for analyzing the nature of invertible transformations developed in this paper may be extensible to noninvertible transformations.

In light of this situation, we show the following theorem on invertible transformations:
If two frames\footnote{Although the use of the word ``system'' would be more appropriate than ``frame,'' we use the latter in connection with disformal transformations.} are related by a general invertible transformation, the EL equations in the new frame are completely equivalent to the original-frame EL equations written in terms of the new fields.
In other words, the new-frame EL equations are derived from the original-frame EL equations without any loss/gain of information of the equations, and vice versa.
Combining this result with the property of invertible transformations that the fields in the two frames are related by a one-to-one correspondence, it can be concluded that any solution of the EL equations in the original frame is mapped to a solution in the new frame by the invertible transformation.
The application of the theorem is not restricted to scalar-tensor theories, but rather extends to any field theory.

This paper is organized as follows.
In \S \ref{sec:toy}, we provide two examples to illustrate the role of (derivative-dependent) invertible transformations.
Then in \S \ref{sec:mth}, we prove the main theorem to clarify the relation between invertible transformations and EL equations.
Furthermore, we present applications of our theorem to scalar-tensor theories in \S \ref{sec:ST}, which include the class of disformal transformations mentioned above.
Finally, we draw our conclusions in \S \ref{sec:conclusion}.

%%%%%%%%%%%%%%%%%%%%%%%%%%%%%%%%%%%%%%%%%%%%%%%%%%%%%%%%%%%%%%%%%%%%%%%%%%%%%%%%%%%%
%%%%%%%%%%%%%%%%%%%%%%%%%%%%%%%%%%%%%%%%%%%%%%%%%%%%%%%%%%%%%%%%%%%%%%%%%%%%%%%%%%%%
%	Examples
%%%%%%%%%%%%%%%%%%%%%%%%%%%%%%%%%%%%%%%%%%%%%%%%%%%%%%%%%%%%%%%%%%%%%%%%%%%%%%%%%%%%
%%%%%%%%%%%%%%%%%%%%%%%%%%%%%%%%%%%%%%%%%%%%%%%%%%%%%%%%%%%%%%%%%%%%%%%%%%%%%%%%%%%%
\section{Examples}\label{sec:toy}

Before proceeding to general arguments in field theories, we give two examples which are useful to get a flavor of the main theorem.

%%%%%%%%%%%%%%%%%%%%%%%%%%%%%%%%%%%%%%%%%%
%%%%%%%%%%%%%%%%%%%%%%%%%%%%%%%%%%%%%%%%%%
\subsection{Analytical mechanics}\label{ssec:ex1}

First, we consider a simple model in analytical mechanics.
Let us start from the Lagrangian
	\be
	L(\dot{X},\dot{Y})=\fr{1}{2}\dot{X}^2+\fr{1}{2}\dot{Y}^2. \label{toylag1}
	\ee
As is obvious, the equations of motion~(EOMs) obtained from this Lagrangian
	\be
	\mE_X\equiv-\ddot{X}=0,~~~\mE_Y\equiv-\ddot{Y}=0, \label{toyeom1}
	\ee
are a pair of second-order ordinary differential equations, and thus we need four initial conditions, i.e., the system has two DOFs.
Now we perform a derivative-dependent frame transformation with
	\be
	X=x-\dot{y},~~~Y=y. \label{toytrnsf1}
	\ee
Note that this transformation is invertible: It can be uniquely solved for $x,y$ as
	\be
	x=X+\dot{Y},~~~y=Y. \label{toytrnsf2}
	\ee
Since $X$ has $\dot y$ in its transformation rule, the new Lagrangian contains a higher-order time derivative:
	\be
	L'(\dot{x},\dot{y},\ddot{y})=\fr{1}{2}(\dot{x}-\ddot{y})^2+\fr{1}{2}\dot{y}^2, \label{toylag2}
	\ee
and so do the EOMs:
	\be
	\mE_x\equiv-\ddot{x}+y^{(3)}=0,~~~\mE_y\equiv-\ddot{y}-x^{(3)}+y^{(4)}=0. \label{toyeom2}
	\ee
At a first glance, this new system of equations seems to require more initial conditions than Eq.~\eqref{toyeom1}, but this is not true.
Indeed, one can eliminate the higher derivative terms by taking linear combinations of the EOMs together with their time derivatives:
	\be
	\begin{split}
	\mE_x+\dot{\mE}_y-\ddot{\mE}_x&=-\ddot{x}=0, \\
	\mE_y-\dot{\mE}_x&=-\ddot{y}=0.
	\end{split}
	\ee
This system of equations has the same structure as the original one~\eqref{toyeom1}.
Therefore, we need the same number of initial conditions to fix the dynamics of $x,y$ as in Eq.~\eqref{toyeom1}.
The above equivalence between the two frames can also be understood as follows.
Written in terms of the original set of variables $(X,Y)$, the left-hand sides of Eq.~\eqref{toyeom2} become
	\be
	\mE_x=-\ddot{X},~~~\mE_y=-\ddot{Y}-X^{(3)}.
	\ee
Then, they are combined to give $\mE_X$ and $\mE_Y$, i.e., the original set of EOMs~\eqref{toyeom1}, as
	\be
	\mE_X=\mE_x,~~~\mE_Y=\mE_y-\dot{\mE}_x, \label{reltoyeom1}
	\ee
while $(\mE_x,\mE_y)$ is expressed in terms of $(\mE_X,\mE_Y)$ as
	\be
	\mE_x=\mE_X,~~~\mE_y=\mE_Y+\dot{\mE}_X. \label{reltoyeom2}
	\ee
Equations~\eqref{reltoyeom1} and \eqref{reltoyeom2} imply that the new-frame EOMs written in terms of the old variables are completely equivalent to the old-frame EOMs.
Hence, any solution in the old frame~$(X,Y)$ is mapped to a solution in the new frame~$(x,y)$ and vice versa, meaning that the two theories~\eqref{toylag1} and \eqref{toylag2} have a common number of physical DOFs.
Note also the similarity between Eqs.~\eqref{toytrnsf1}, \eqref{toytrnsf2} and Eqs.~\eqref{reltoyeom1}, \eqref{reltoyeom2}.
We shall clarify the origin of the similarity in \S \ref{sec:mth}. 

One may notice that the transformation~\eqref{toytrnsf1} basically captures the essential nature of the disformal transformation~\eqref{disformal}. 
The crucial difference between them is that the disformal transformation is more complicated so that it is not always invertible.
When it is invertible, the logic is the same as the above discussion.

%%%%%%%%%%%%%%%%%%%%%%%%%%%%%%%%%%%%%%%%%%
%%%%%%%%%%%%%%%%%%%%%%%%%%%%%%%%%%%%%%%%%%
\subsection{Scalar-tensor theory}\label{ssec:ex2}

The second example is the case of scalar-tensor theory.
For the Einstein-Hilbert action with a canonical scalar field~$\ti{\phi}$ and some matter fields~$\Psi^I$,
	\be
	S[\ti{g}_{\mu\nu},\ti{\phi};\Psi^I]=\int d^4x\sqrt{-\ti{g}}\brb{\fr{\Mpl^2}{2}\ti{R}-\fr{1}{2}\ti{\na}_\mu\ti{\phi}\ti{\na}^\mu\ti{\phi}-V(\ti{\phi})}+S_{\rm m}[\ti{g}_{\mu\nu};\Psi^I], \label{grscalar}
	\ee
let us consider the following transformation:
	\be
	\ti{g}_{\mu\nu}=g_{\mu\nu},~~~\ti{\phi}=\phi-f(R), \label{f(R)}
	\ee
where $f(R)$ is an arbitrary function of the Ricci scalar associated with the metric $g_{\mu\nu}$.
The inverse transformation is given by
	\be
	g_{\mu\nu}=\ti{g}_{\mu\nu},~~~\phi=\ti{\phi}+f(\ti{R}),
	\ee
where now $\ti{R}$ is computed from $\ti{g}_{\mu\nu}$.
For this transformation, the original action~\eqref{grscalar} is transformed as
	\be
	S'[g_{\mu\nu},\phi;\Psi^I]=\int d^4x\sqrt{-g}\brb{\fr{\Mpl^2}{2}R-\fr{1}{2}\na_\mu\bra{\phi-f(R)}\na^\mu\bra{\phi-f(R)}-V\!\bra{\phi-f(R)}}+S_{\rm m}[g_{\mu\nu};\Psi^I].
	\ee
Introducing a Lagrange multiplier, one can recast this $S'$ into the form of
	\be
	S''[g_{\mu\nu},\phi,\chi,\la;\Psi^I]=\int d^4x\sqrt{-g}\brb{\fr{\Mpl^2}{2}R-\fr{1}{2}\na_\mu\varphi\na^\mu\varphi-V(\varphi)+\la(\chi-R)}+S_{\rm m}[g_{\mu\nu};\Psi^I],~~~\varphi\equiv \phi-f(\chi), \label{trigal}
	\ee
which manifestly yields second-order field equations.
The action \eqref{trigal}, which contains three scalar fields $\phi$, $\chi$, and $\lambda$, describes a specific model of tensor-multiscalar theory defined in Ref.~\cite{Damour:1992we}.
Although such a theory has $2+3$ DOFs in general, the specific theory defined by $S''$ is expected to have only $2+1$ DOFs as it is obtained via the invertible transformation~\eqref{f(R)} from the action \eqref{grscalar} containing only one scalar field.
Actually, we can explicitly show that the EL equations derived from $S''$ are completely equivalent to those derived from the original action~$S$ in the following way.
The EOMs obtained from $S''$ are
	\be
	E_{\mu\nu}\equiv\fr{1}{\sqrt{-g}}\fr{\delta S''}{\delta g^{\mu\nu}}=0,~~~E_\Phi\equiv\fr{1}{\sqrt{-g}}\fr{\delta S''}{\delta \Phi}=0,~~~(\Phi=\phi,\chi,\la), \label{trigaleom}
	\ee
where
	\begin{align}
	E_{\mu\nu}&=\fr{\Mpl^2}{2}\bra{R_{\mu\nu}-\fr{1}{2}Rg_{\mu\nu}}+\fr{1}{2}g_{\mu\nu}\brb{\fr{1}{2}\na_\si\varphi\na^\si\varphi+V(\varphi)-\la(\chi-R)} \nonumber \\
	&\quad -\fr{1}{2}\na_\mu\varphi\na_\nu\varphi+\na_\mu\na_\nu\la-g_{\mu\nu}\Box\la-\la R_{\mu\nu}-\fr{1}{2}T_{\mu\nu}, \\
	E_{\phi}&=\Box\varphi-V'(\varphi), \\
	E_{\chi}&=\la-f'(\chi)\brb{\Box\varphi-V'(\varphi)}, \\
	E_{\la}&=\chi-R,
	\end{align}
with $T_{\mu\nu}\equiv-\fr{2}{\sqrt{-g}}\fr{\delta S_{\rm m}}{\delta g^{\mu\nu}}$ being the energy-momentum tensor for the matter fields.
The EOM for the Lagrange multiplier~$E_{\la}=0$ implies $\chi=R$.
Combining $E_{\phi}=0$ and $E_{\chi}=0$, one obtains $\la=0$.
Thus, the metric EOM~$E_{\mu\nu}=0$ is written as
	\be
	\fr{\Mpl^2}{2}\bra{R_{\mu\nu}-\fr{1}{2}Rg_{\mu\nu}}+\fr{1}{2}g_{\mu\nu}\brb{\fr{1}{2}\na_\si\varphi\na^\si\varphi+V(\varphi)}-\fr{1}{2}\na_\mu\varphi\na_\nu\varphi-\fr{1}{2}T_{\mu\nu}=0. \label{Einstein}
	\ee
This equation and $E_{\phi}=0$ are nothing but the Einstein and Klein-Gordon equations derived from the original action~\eqref{grscalar} with the replacement~$\phi\to\varphi$.
Hence, for any solution~$(g_{\mu\nu},\phi)=(g_{\mu\nu}^{(0)},\phi^{(0)})$ of the original EOMs, the set~$(g_{\mu\nu},\varphi)=(g_{\mu\nu}^{(0)},\phi^{(0)})$ satisfies the new-frame EOMs.
Once $(g_{\mu\nu},\varphi)$ is fixed, the solution for the new-frame variables~$(g_{\mu\nu},\phi,\chi,\la)$ is also fixed as follows:
	\be
	g_{\mu\nu}=g_{\mu\nu}^{(0)},~~~\phi=\phi^{(0)}+f(R^{(0)}),~~~\chi=R^{(0)},~~~\la=0,
	\ee
where $R^{(0)}$ denotes the Ricci scalar associated with $g_{\mu\nu}^{(0)}$.
Therefore, the system of EOMs~\eqref{trigaleom} is essentially the system composed of the Einstein equation~\eqref{Einstein} and the Klein-Gordon equation~$E_{\phi}=0$, and thus has the same number of DOFs as the original system.
\vskip0.3cm

What we can learn from these simple examples is that, even if we perform a derivative-dependent transformation to obtain a Lagrangian with higher derivatives, it has the same number of DOFs as the original one as long as the transformation is invertible.
In the subsequent section, we prove this statement for general field theories.

%%%%%%%%%%%%%%%%%%%%%%%%%%%%%%%%%%%%%%%%%%%%%%%%%%%%%%%%%%%%%%%%%%%%%%%%%%%%%%%%%%%%
%%%%%%%%%%%%%%%%%%%%%%%%%%%%%%%%%%%%%%%%%%%%%%%%%%%%%%%%%%%%%%%%%%%%%%%%%%%%%%%%%%%%
%	Main theorem
%%%%%%%%%%%%%%%%%%%%%%%%%%%%%%%%%%%%%%%%%%%%%%%%%%%%%%%%%%%%%%%%%%%%%%%%%%%%%%%%%%%%
%%%%%%%%%%%%%%%%%%%%%%%%%%%%%%%%%%%%%%%%%%%%%%%%%%%%%%%%%%%%%%%%%%%%%%%%%%%%%%%%%%%%
\section{Proof of the theorem}\label{sec:mth}

%%%%%%%%%%%%%%%%%%%%%%%%%%%%%%%%%%%%%%%%%%
%%%%%%%%%%%%%%%%%%%%%%%%%%%%%%%%%%%%%%%%%%
\subsection{Setup}\label{ssec:setup}

Let us consider a general field theory in $D$-dimensional spacetime:
	\be
	\begin{split}
	S&=\int d^Dx\, L[\phi],\\
	L[\phi]&\equiv L(\phi^i,\pa_\mu\phi^i,\pa_\mu\pa_\nu\phi^i,\cdots,\pa_{(n)}\phi^i),
	\end{split} \label{gft}
	\ee
where $i=1,\cdots,N$ labels the fields and $\pa_{(k)}\equiv\pa_{\mu_1}\cdots\pa_{\mu_k}$.
Transforming $\phi^i$ to a new set of fields $\psi^i$ by\footnote{Although it is natural to begin with the expression of the new variables in terms of the old ones, i.e., in the form of $\psi^i=g^i[\phi]$, we instead start from Eq.~\eqref{tr} which is more convenient for later arguments.}
	\be
	\phi^i=f^i[\psi]\equiv f^i(\psi^j,\pa_\mu\psi^j,\pa_\mu\pa_\nu\psi^j,\cdots,\pa_{(m)}\psi^j), \label{tr}
	\ee
we obtain a new theory, symbolically written as
	\be
	L'[\psi]\equiv L[f[\psi]],
	\ee
which consists of at most $(m+n)$th derivatives of $\psi^i$.
It should be noted that the transformation~\eqref{tr} depends only on $\psi^i$ and their derivatives evaluated at the same point in the spacetime.
A field transformation between $\phi^i$ and $\psi^i$ is called {\it invertible} if $\psi^i$ are uniquely determined from $\phi^i$ and vice versa.
As such, Eq.~\eqref{tr} can be solved for $\psi^i$ in the form of
	\be
	\psi^i=g^i(\phi^j,\pa_\mu\phi^j,\pa_\mu\pa_\nu\phi^j,\cdots,\pa_{(\ell)}\phi^j). \label{invtr}
	\ee
Hereafter we require that the number of $\phi$ fields be the same as that of $\psi$ fields, because otherwise one cannot define an invertible transformation between $\phi^i$ and $\psi^i$.\footnote{We also require that the dynamics of $\phi^i$ is restricted within the codomain of $f^i$.}

In general, the transformation law~\eqref{tr} could be quite nonlinear.
However, as we shall see below, if the invertibility is considered only locally in field space, the invertibility of the transformation can be judged within the language of linear algebra.
Let us consider infinitesimal changes~$\delta\phi^i, \delta\psi^i$ from configurations of $\phi^i, \psi^i$ that satisfy the relation~\eqref{tr}.
Then Eq.~\eqref{tr} is linearized as
	\be
	\delta\phi^i=\hat{P}^i_j\delta\psi^j, \label{inftr}
	\ee
where $\hat{P}^i_j$ is a derivative-operator-valued matrix determined from the functional form of $f^i$:
	\be
	\hat{P}^i_j=\sum_{s=0}^mu^{i(s)}_j\pa_{(s)},~~~u^{i(s)}_j\equiv\fr{\pa f^i}{\pa(\pa_{(s)}\psi^j)}.
	\ee
A system of equations of the form~\eqref{inftr} is called linear differential-algebraic equations (DAEs), since it consists of coupled linear differential and algebraic equations.
The solution to Eq.~\eqref{inftr} is generically not unique as it may contain integration constants.
On the other hand, if the transformation~\eqref{tr} is invertible at least locally, then one can uniquely solve the system of DAEs~\eqref{inftr} for $\delta\psi^i$ in the form of
	\be
	\delta\psi^i=\hat{Q}^i_j\delta\phi^j, \label{inftr2}
	\ee
where $\hat{Q}^i_j$ is a derivative-operator-valued matrix satisfying\footnote{If there exists a derivative-operator-valued matrix $\hat{Q}^i_j$ for which $\hat{Q}^i_j\hat{P}^j_k=\delta^i_k$, one can prove $\hat{P}^i_j\hat{Q}^j_k=\delta^i_k$ and the uniqueness of such $\hat{Q}^i_j$ in the same manner as $c$-number matrices.}
	\be
	\hat{P}^i_j\hat{Q}^j_k=\hat{Q}^i_j\hat{P}^j_k=\delta^i_k, \label{invopr}
	\ee
and hence plays the role of the inverse operator of $\hat{P}^i_j$.
In the present paper, we restrict ourselves to such a special class of field transformations.
For the detailed arguments on the unique solvability of DAEs, see Ref.~\cite{Motohashi:2016prk}.

%%%%%%%%%%%%%%%%%%%%%%%%%%%%%%%%%%%%%%%%%%
%%%%%%%%%%%%%%%%%%%%%%%%%%%%%%%%%%%%%%%%%%
\subsection{Main theorem}\label{ssec:main}

In \S \ref{sec:toy}, we saw that an invertible transformation does not change the number of physical DOFs in two simple models.
Below we prove the following theorem for general field theories:

\vskip0.3cm
{\bf Theorem.} {\it
Suppose two sets of fields $\phi^i$ and $\psi^i$ are related by an invertible transformation of the form~$\phi^i=f^i[\psi]$.
If a configuration~$\psi^i_{(0)}$ satisfies the EL equations for $\psi^i$, then its transformation~$f^i[\psi_{(0)}]$ satisfies the EL equations for $\phi^i$.
Conversely, if a configuration $\phi^i_{(0)}$ satisfies the EL equations for $\phi^i$, then its inverse transformation~$(f^{-1})^i[\phi_{(0)}]$ satisfies the EL equations for $\psi^i$.\footnote{The symbol~$(f^{-1})^i[\phi]$ stands for the configuration of $\psi^i$ that satisfies $\phi^i=f^i[\psi]$.}
}
\vskip0.3cm

{\it Proof.}
Let us consider the variation of the action in two different manners.
If we vary the original action written in terms of $\phi^i$, then we obtain
	\be
	\delta S=\delta\int d^Dx\, L[\phi]=\int d^Dx\, \mE_i^{(\phi)}\delta\phi^i. \label{varwrtphi}
	\ee
Here, $\mE_i^{(\phi)}=0$ denote the EOMs for $\phi^i$.
Meanwhile, if we rewrite the action in terms of $\psi^i$ by the relation~\eqref{tr}, the variation yields
	\be
	\delta S=\delta\int d^Dx\, L'[\psi]=\int d^Dx\, \mE_i^{(\psi)}\delta\psi^i, \label{varwrtpsi}
	\ee
where $\mE_i^{(\psi)}=0$ are the EOMs for $\psi^i$.
Note that, in deriving the EL equations, we have imposed independent boundary conditions for $\phi^i$ and $\psi^i$:
$\pa_{(k)}\phi^i=0\,(k=0,1,\cdots,n-1)$ and $\pa_{(k')}\psi^i=0\,(k'=0,1,\cdots,n+m-1)$, respectively.
This is because we need only the relation between the old- and new-frame EL equations obtained in such manner.
Now we impose Eq.~\eqref{inftr} on $\delta\phi^i$ and reexpress Eq.~\eqref{varwrtphi} by $\delta\psi^i$:\footnote{Hereafter $\phi^i$ and $\psi^i$ are freely replaced with each other via the relation~\eqref{tr}.}
	\be
	\delta S=\int d^Dx\, \mE_i^{(\phi)}\hat{P}^i_j\delta\psi^j=\int d^Dx\bra{\hat{P}^\da{}^i_j\mE_i^{(\phi)}}\delta\psi^j. \label{varwrtphi2}
	\ee
Note that, as a result of integration by parts, here we have the adjoint of $\hat{P}^i_j$ which satisfies
	\be
	\hat{P}^\da{}^i_jw_i=\sum_{s=0}^m(-1)^s\pa_{(s)}\bra{u^{i(s)}_jw_i},
	\ee
with $w_i$ being an arbitrary vector function.
After this, one can compare Eqs.~\eqref{varwrtpsi} and \eqref{varwrtphi2}.
Since $\delta\psi^i$ are arbitrary, one obtains the following relation between $\mE_i^{(\phi)}$ and $\mE_i^{(\psi)}$:\footnote{The relation~\eqref{eomrel} itself holds even if the transformation is not invertible.}
	\be
	\mE_i^{(\psi)}=\hat{P}^\da{}^j_i\mE_j^{(\phi)}. \label{eomrel}
	\ee

The relation \eqref{eomrel} can be regarded as the adjoint DAE system to Eq.~\eqref{inftr}.
Now the problem is whether the original set of EOMs~$\mE_i^{(\phi)}=0$ follows from $\hat{P}^\da{}^j_i\mE_j^{(\phi)}=0$.
To prove this, one can follow the arguments on the unique solvability of adjoint DAEs given in Ref.~\cite{Motohashi:2016prk}.
Since the operator matrix~$\hat{P}^i_j$ has its inverse~$\hat{Q}^i_j$ without integral operator, one can take the adjoint of Eq.~\eqref{invopr} to obtain
	\be
	\hat{P}^\da{}^i_j\hat{Q}^\da{}^j_k=\hat{Q}^\da{}^i_j\hat{P}^\da{}^j_k=\delta^i_k,
	\ee
which means that the inverse operator of $\hat{P}^\da{}^i_j$ is independent of integral operators and given by $\hat{Q}^\da{}^i_j$.  
Therefore, from Eq.~\eqref{eomrel} we obtain 
	\be
	\mE_i^{(\phi)}=\hat{Q}^\da{}^j_i\mE_j^{(\psi)}, \label{eomrel2}
	\ee
which is the adjoint DAE system to Eq.~\eqref{inftr2}. 
By multiplying both sides of $\hat{P}^\da{}^j_i\mE_j^{(\phi)}=0$ by $\hat{Q}^\da{}^i_k$ yields $\mE_k
^{(\phi)}=0$.
Hence, if a configuration of $\psi^i$ that satisfies the new set of EOMs~\eqref{eompsi} is transformed by the relation~\eqref{tr}, then the resulting configuration of $\phi^i$ satisfies the original set of EOMs~\eqref{eomphi}.
Moreover, the opposite direction is also true.
This completes the proof of the main theorem.
\hfill $\Box$
\vskip0.3cm

In deriving the relation~\eqref{eomrel}, we have not used explicit expressions for the EL equations in each frame.
Although technically more complicated, it is also possible to show the relation by a direct comparison between the explicit expression of $\mE_i^{(\phi)}$ and that of $\mE_i^{(\psi)}$ as follows.
The EOMs for $\phi^i$ are formally written as
	\be
	\mE_i^{(\phi)}\equiv\fr{\delta L[\phi]}{\delta\phi^i}=\sum_{q=0}^n(-1)^q\pa_{(q)}v_i^{(q)}=0,~~~v_i^{(q)}\equiv\fr{\pa L}{\pa(\pa_{(q)}\phi^i)}. \label{eomphi}
	\ee
On the other hand, the EOMs for $\psi^i$ become\footnote{Note that $\mE_i^{(\psi)}=0$ is different from what one obtains by substituting Eq.~\eqref{tr} into $\mE_i^{(\phi)}=0$, though these two sets of equations are equivalent by virtue of the relation~\eqref{eomrel}.}
	\be
	\mE_i^{(\psi)}\equiv\fr{\delta L[f[\psi]]}{\delta\psi^i}=\sum_{p=0}^{m+n}\sum_{q=0}^n(-1)^p\pa_{(p)}\brb{v_j^{(q)}\fr{\pa(\pa_{(q)}f^j)}{\pa(\pa_{(p)}\psi^i)}}=0. \label{eompsi}
	\ee
By using the relation\footnote{Equation~\eqref{ddfddpsi} can be verified by repeated use of the following identity for $\Phi[\psi]=\pa_{(r)}f^j[\psi]\,(r=0,1,\cdots,q-1)$:
	\be
	\fr{\pa(\pa_{(1)}\Phi[\psi])}{\pa(\pa_{(p)}\psi^i)}=\pa_{(1)}\fr{\pa\Phi[\psi]}{\pa(\pa_{(p)}\psi^i)}+\fr{\pa\Phi[\psi]}{\pa(\pa_{(p-1)}\psi^i)}, \nonumber
	\ee
which can be checked by expanding the both sides using the chain rule.}
	\be
	\fr{\pa(\pa_{(q)}f^j)}{\pa(\pa_{(p)}\psi^i)}=\sum_{\substack{0\le k\le q \\ 0\le p-k\le m}}\begin{pmatrix}q\\k\end{pmatrix}\pa_{(q-k)}u^{j(p-k)}_i, \label{ddfddpsi}
	\ee
the expression of $\mE_i^{(\psi)}$ becomes
	\be
	\mE_i^{(\psi)}=\sum_{k=0}^n\sum_{p=k}^{k+m}\sum_{q=k}^n(-1)^p\begin{pmatrix}q\\k\end{pmatrix}\pa_{(p)}\brb{v_j^{(q)}\pa_{(q-k)}u^{j(p-k)}},
	\ee
where we have interchanged the summations.
With the aid of the Leibniz rule,
	\begin{align}
	\mE_i^{(\psi)}&=\sum_{k=0}^n\sum_{p=k}^{k+m}\sum_{q=k}^n\sum_{r=0}^k(-1)^p \begin{pmatrix}q\\k\end{pmatrix}\begin{pmatrix}k\\r\end{pmatrix}\pa_{(p-k)}\brb{\pa_{(r)}v_j^{(q)}\pa_{(q-r)}u^{j(p-k)}} \nonumber \\
	&=\sum_{s=0}^m\sum_{k=0}^n\sum_{q=k}^n\sum_{r=0}^k(-1)^{k+s} \begin{pmatrix}q\\k\end{pmatrix}\begin{pmatrix}k\\r\end{pmatrix}\pa_{(s)}\brb{\pa_{(r)}v_j^{(q)}\pa_{(q-r)}u^{j(s)}},
	\end{align}
where we have defined $s\equiv p-k$.
Interchanging the summations as $\sum_{k=0}^n\sum_{q=k}^n\sum_{r=0}^k=\sum_{q=0}^n\sum_{r=0}^q\sum_{k=r}^q$ and using the formula
	\be
	\sum_{k=r}^q(-1)^k\begin{pmatrix}q\\k\end{pmatrix}\begin{pmatrix}k\\r\end{pmatrix}=(-1)^q\delta_{qr},
	\ee
we finally obtain
	\be
	\mE_i^{(\psi)}=\sum_{s=0}^m\sum_{q=0}^n(-1)^{s+q}\pa_{(s)}\brb{u^{j(s)}_i\pa_{(q)}v_j^{(q)}}=\hat{P}^\da{}^j_i\mE_j^{(\phi)},
	\ee
which is nothing but Eq.~\eqref{eomrel}.

%%%%%%%%%%%%%%%%%%%%%%%%%%%%%%%%%%%%%%%%%%
%%%%%%%%%%%%%%%%%%%%%%%%%%%%%%%%%%%%%%%%%%
\subsection{Remarks}\label{ssec:remarks}

According to the Theorem, if one can define an inverse transformation between $\phi^i$ and $\psi^i$, then the solution space for $\psi^i$ is mapped to a subspace of the solution space for $\phi^i$, and vice versa.
Therefore, the two solution spaces have the same number of DOFs.

One may naively think that the proof for the equivalence between $\mE_i^{(\phi)}=0$ and $\hat{P}^\da{}^i_j\mE_i^{(\phi)}=0$ becomes simpler if the field transformation is reduced to one without field derivatives.
Such a reduction is realized by introducing auxiliary fields with Lagrange multipliers and then replacing the derivatives contained in the field transformation by the auxiliary fields.
However, this method is not helpful for comparing $\mE_i^{(\phi)}=0$ and $\hat{P}^\da{}^i_j\mE_i^{(\phi)}=0$ as the EOMs of the resulting theory apparently do not coincide with $\hat{P}^\da{}^i_j\mE_i^{(\phi)}=0$.
For the detailed arguments, see the Appendix.

There are several other remarks on the Theorem.
If a given transformation law~$\phi^i=f^i[\psi]$ can be solved for $\psi^i$ without integration constant but with branches of solutions, one can still apply the Theorem by choosing any one of the branches. 
For instance, the transformation $\phi=\psi^2$ has two inverse transformations, $\psi=\pm\sqrt{\phi}$.
In this case, $\hat P = 2\psi$ and $\hat Q = \pm 1/(2\sqrt{\phi})$.
After choosing either of the branches of $\hat Q$, one can apply the Theorem.

The original- and new-frame EL equations~($\mE_i^{(\phi)}=0$ and $\mE_i^{(\psi)}=0$) are at most $(2n)$th- and $(2n+2m)$th-order differential equations, respectively.
We have seen in Eq.~\eqref{eomrel2} that the latter can be reduced to the equivalent $(2n+m)$th-order differential equations by operating the regular matrix~$\hat{Q}^\da{}^i_j$.
Since the Theorem states that the DOFs of the reduced equations are the same as the ones of the original-frame EL equations, it is natural to expect that the reduced equations can be further reduced to the manifestly equivalent $(2n)$th-order differential equations by some manipulations.
While the total number of required initial conditions is the same in both frames, it does not mean EL equations in the new frame are reducible to the set of equations each of which has the same orders of derivatives as those for EL equations in the old frame.
To see this, let us consider the following Lagrangian in analytical mechanics depending on $(X,Y)$:
	\be
	L=\fr{1}{2}\dot{X}^2+\fr{1}{2}\dot{Y}^2+XY. \label{irreducible}
	\ee
Clearly, the EOMs for $(X,Y)$,
	\be
	\mE_X \equiv -\ddot{X} + Y =0,~~~\mE_Y \equiv -\ddot{Y} + X =0 , \label{EXY}
	\ee
are two second-order differential equations, which require four initial conditions.
Now we perform a transformation of variables~$(X,Y)\to (x,y)$ defined by
	\be
	x=\mE_Y = X - \ddot{Y} ,~~~y=Y, \label{trnsf2}
	\ee
which has the inverse transformation
	\be
	X = x + \ddot{y},~~~Y = y . \label{invtrnsf2}
	\ee
Note that $x=0$ and is nondynamical by definition.
The Lagrangian is then transformed as
	\be
	L' = \fr{1}{2} (\dot{x} + y^{(3)})^2 - \fr{1}{2} \dot{y} ^2 + xy ,
	\ee
where we performed integration by parts.
The EOMs for $(x,y)$ are given by
	\be
	\mE_x \equiv -\ddot{x} - y^{(4)} + y = 0,~~~\mE_y \equiv - x^{(4)} - y^{(6)} + \ddot{y} + x = 0 .  \label{Exy}
	\ee
Due to our main theorem, the EOMs in the new frame are equivalent to those in the old frame through the relation~\eqref{eomrel2} with the replacement~\eqref{invtrnsf2}.
Indeed,  
	\be
	\begin{split}  
	\mE_X &= \mE_x = - \ddot{x} - y^{(4)} + y = 0 ,\\
	\mE_Y &= \mE_y - \ddot{\mE}_x = x = 0 .
	\end{split}
	\ee
Hence, the EOMs for $(x,y)$ are $x=0$, which is consistent with Eq.~\eqref{trnsf2}, and a fourth-order differential equation~$-y^{(4)}+y=0$ obtained by substituting $x=0$ into $\mE_X=0$.
As such, the two second-order equations~\eqref{EXY} are transformed into one zeroth-order equation and one fourth-order equation.
Obviously, the new EOMs are not reducible to two second-order differential equations. 
Nevertheless, the two sets are still related through Eq.~\eqref{eomrel} or Eq.~\eqref{eomrel2}, and have the same number of DOFs.
In this case, the EOMs in both frames indeed require four initial conditions.
This example demonstrates that in general the old and new sets of EOMs have different derivative structures.\footnote{This type of situation happens whenever one defines an invertible transformation in such a way that a part of the new fields becomes nondynamical. The above example has a problem that the Hamiltonian obtained from the Lagrangian~\eqref{irreducible} is not bounded below. Such a model was chosen just for simplicity in calculation.}
To clarify their structures, one has to investigate on a case-by-case basis.

Another thing to note is that if the original theory has gauge symmetries, then Eq.~\eqref{eomrel} is not the only way to express $\mE_i^{(\psi)}$ in terms of $\mE_i^{(\phi)}$.
This is because there exist identities among the EL equations corresponding to the gauge symmetries, i.e., Noether identities.
If the original theory is invariant under an infinitesimal gauge transformation in the form of
	\be
	\Delta_\ep\phi^i=\hat{G}^i_I\ep^I,
	\ee
where $I=1,\cdots,M$ labels the gauge symmetries, then the Noether identities are written as\footnote{The number of the gauge symmetries~$M$ is smaller than that of the fields~$N$.}
	\be
	\hat{G}^\da{}^i_I\mE_i^{(\phi)}=0, \label{Noether}
	\ee
which reduces the dimensionality of the old-frame EOM space by $M$.
Correspondingly, the new system also has gauge symmetries under the infinitesimal transformation~$\psi^i\to\psi^i+\Delta_\ep\psi^i$, where $\Delta_\ep\psi^i=\hat{Q}^i_j\hat{G}^j_I\ep^I$.
Therefore, the new-frame EOMs satisfy the corresponding Noether identities~$\hat{G}^\da{}^i_I\hat{Q}^\da{}^j_i\mE_j^{(\psi)}=0$, which means that the new-frame EOM space also has dimension $N-M$.
Even in this case, the proof of the main theorem still holds since it relies only on the invertibility of $\hat{P}^\da{}^i_j$.
On the other hand, combining Eqs.~\eqref{eomrel} and \eqref{Noether}, we obtain
	\be
	\mE_i^{(\psi)}=(\hat{P}^\da{}^j_i+\hat{F}^I_i\hat{G}^\da{}^j_I)\mE_j^{(\phi)}\equiv \hat{R}^j_i\mE_j^{(\phi)},
	\ee
with $\hat{F}^I_i$ being an arbitrary derivative-operator-valued matrix.
Note that this arbitrariness of the relation between the EOMs does not spoil the proof of the main theorem.
For some choice of $\hat{F}^I_i$, the matrix~$\hat{R}^i_j$ may become singular, in which case $\hat{R}^i_j$ is a projection operator onto the $(N-M)$-dimensional constrained surface in the $N$-dimensional EOM space defined by the Noether identity~\eqref{Noether}.
Nevertheless, the singularity is not problematic since it is only this constrained surface that is physically relevant.

Before closing this section, let us remark that not all the variables relevant to an invertible transformation have to be dynamical, in which case however the transformed theory acquires redundant DOFs in general.
We consider the following Lagrangian as an example:
	\be
	L(\dot{X})=\fr{1}{2}\dot{X}^2, \label{toylag3}
	\ee
with the invertible field transformation of the same form as Eq.~\eqref{toytrnsf1}.
Note that $Y$ does not appear in the original Lagrangian~\eqref{toylag3}.
In this case, the new Lagrangian takes the form
	\be
	L'(\dot{x},\ddot{y})=\fr{1}{2}(\dot{x}-\ddot{y})^2,
	\ee
which has a gauge symmetry under
	\be
	x\to x+\dot{\xi},~~~y\to y+\xi,
	\ee
with $\xi$ being an arbitrary function of time.
Once the gauge is completely fixed by setting $y=0$, we recover the original Lagrangian.
In other words, introducing a gauge DOF~$y$ to the original theory defined by $L$ is an invertible transformation, whose inverse is fixing the gauge completely by setting $y=0$ in the resultant new theory described by $L'$, and vice versa.

This example is related to the St\"{u}ckelberg formalism for a massive vector field.
We start from the Proca Lagrangian
	\be
	L_{\rm Proca}(\ti{A}_\mu,\pa_\la\ti{A}_\mu)=-\fr{1}{4}\ti{F}_{\mu\nu}\ti{F}^{\mu\nu}+m^2\ti{A}_\mu \ti{A}^\mu,~~~
	\ti{F}_{\mu\nu}\equiv \pa_\mu\ti{A}_\nu-\pa_\nu\ti{A}_\mu.
	\ee
One can restore $U(1)$ gauge symmetry via introducing a St\"{u}ckelberg scalar $\phi$ by promoting
	\be
	\ti{A}_\mu\to A_\mu-\pa_\mu\phi, \label{procast}
	\ee
and assuming the following gauge transformation law
	\be
	A_\mu\to A_\mu+\pa_\mu\Lambda,~~~\phi\to \phi+\Lambda. \label{restU1}
	\ee
Indeed, the new Lagrangian
	\be
	L'_{\rm Proca}(A_\mu,\phi,\pa_\la A_\mu,\pa_\la\phi)=-\fr{1}{4}F_{\mu\nu}F^{\mu\nu}+m^2(A_\mu-\pa_\mu\phi)(A^\mu-\pa^\mu\phi),~~~
	F_{\mu\nu}\equiv \pa_\mu A_\nu-\pa_\nu A_\mu \label{newProca}
	\ee
is invariant under the transformation~\eqref{restU1}.
In this case, the replacement~\eqref{procast} can be regarded as an invertible transformation by identifying it as the following field redefinition:
	\be
	\ti{A}_\mu=A_\mu-\pa_\mu\phi,~~~\ti{\phi}=\phi.
	\ee
The inverse transformation is given by
	\be
	A_\mu=\ti{A}_\mu+\pa_\mu\ti{\phi},~~~\phi=\ti{\phi}. \label{invtrProca}
	\ee
Hence, as it should be, the St\"{u}ckelberg formalism just introduces a redundant DOF and it does not change the number of physical DOFs.
On the other hand, imposing a complete gauge fixing~$\phi=0$ in $L'_{\rm Proca}$ can be identified as performing an invertible transformation~\eqref{invtrProca} on $L'_{\rm Proca}$.

Similarly, in the context of scalar-tensor theories, any additional scalar/vector/tensor fields can be introduced without changing the number of DOFs.
This may be related to the recent work~\cite{Ezquiaga:2017ner}, which suggested a connection between tensor-multiscalar theories~\cite{Damour:1992we}, generalized Proca theories~\cite{Heisenberg:2014rta} and bigravity~\cite{Hassan:2011zd}.

%%%%%%%%%%%%%%%%%%%%%%%%%%%%%%%%%%%%%%%%%%%%%%%%%%%%%%%%%%%%%%%%%%%%%%%%%%%%%%%%%%%%
%%%%%%%%%%%%%%%%%%%%%%%%%%%%%%%%%%%%%%%%%%%%%%%%%%%%%%%%%%%%%%%%%%%%%%%%%%%%%%%%%%%%
%	Scalar-tensor theories
%%%%%%%%%%%%%%%%%%%%%%%%%%%%%%%%%%%%%%%%%%%%%%%%%%%%%%%%%%%%%%%%%%%%%%%%%%%%%%%%%%%%
%%%%%%%%%%%%%%%%%%%%%%%%%%%%%%%%%%%%%%%%%%%%%%%%%%%%%%%%%%%%%%%%%%%%%%%%%%%%%%%%%%%%
\section{Applications to scalar-tensor theories}\label{sec:ST}

In the previous section, we have shown that the new-frame EL equations can be made equivalent to the original-frame EL equations by using the regular matrix~$\hat{P}^\da{}^i_j$.
In this section, we consider two types of invertible transformations in the context of scalar-tensor theories
and present explicit forms of the matrix~$\hat{P}^\da{}^i_j$.

%%%%%%%%%%%%%%%%%%%%%%%%%%%%%%%%%%%%%%%%%%
%%%%%%%%%%%%%%%%%%%%%%%%%%%%%%%%%%%%%%%%%%
\subsection{Disformal transformation}\label{ssec:disformal}

Let us consider the disformal transformation mentioned in \S \ref{sec:introduction}:
	\be
	\ti{g}_{\mu\nu}=A(\phi,X)g_{\mu\nu}+B(\phi,X)\na_\mu\phi\na_\nu\phi,~~~\ti{\phi}=\phi. \label{disformal2}
	\ee
One can define the inverse matrix of $\ti{g}_{\mu\nu}$ as
	\be
	\ti{g}^{\mu\nu}=\fr{1}{A}\bra{g^{\mu\nu}-\fr{B}{A-2XB}\na^\mu\phi\na^\nu\phi},
	\ee
as long as $A(A-2XB)\ne0$.
Note that any composition of disformal transformations is again a disformal transformation.
The necessary and sufficient condition for the invertibility of the disformal transformation is given by~\cite{Zumalacarregui:2013pma}\footnote{The condition $A(A-2XB)\ne0$, which guarantees the existence of the inverse matrix~$\ti{g}^{\mu\nu}$, automatically follows from Eq.~\eqref{disformal-cond}.}
	\be
	A(A-XA_X+2X^2B_X)\ne0, \label{disformal-cond}
	\ee
which ensures the Jacobian determinant for the metric transformation is nonvanishing.
If this is the case, the inverse disformal transformation is written as
	\be
	g_{\mu\nu}=\ti{A}(\ti{\phi},\ti{X})\ti{g}_{\mu\nu}+\ti{B}(\ti{\phi},\ti{X})\ti{\na}_\mu\ti{\phi}\ti{\na}_\nu\ti{\phi},~~~\phi=\ti{\phi}, \label{invdisformal}
	\ee
where $\ti{\na}_\mu$ denotes a covariant derivative with respect to $\ti{g}_{\mu\nu}$, and the canonical kinetic term of the scalar field in the original frame is related to the new variables by
	\be
	\ti{X}\equiv -\fr{1}{2}\ti{g}^{\mu\nu}\ti{\na}_\mu\ti{\phi}\ti{\na}_\nu\ti{\phi}=\fr{X}{A-2XB}. \label{Xrel}
	\ee
The functional forms of $\ti{A},\ti{B}$ are given by the following relation:
	\be
	\ti{A}(\ti{\phi},\ti{X})=\fr{1}{A(\ti{\phi},X)},~~~\ti{B}(\ti{\phi},\ti{X})=-\fr{B(\ti{\phi},X)}{A(\ti{\phi},X)},
	\ee
where $X$ should be written in terms of $(\ti{\phi},\ti{X})$ by solving Eq.~\eqref{Xrel}.
As it should be, the solvability of Eq.~\eqref{Xrel} for $X$ is guaranteed by the condition~\eqref{disformal-cond} as
	\be
	\fr{\pa\ti{X}}{\pa X}=\fr{A-XA_X+2X^2B_X}{(A-2XB)^2}\ne 0.
	\ee

For some known classes of scalar-tensor theories, the transformation properties under disformal transformations have been well studied.
The authors of Ref.~\cite{Bettoni:2013diz} showed that the Horndeski class is closed under disformal transformations with $A,B$ depending on $\phi$ only.
If one proceeds to $X$-dependent~$B$, the Horndeski theories are transformed to GLPV theories~\cite{Gleyzes:2014dya,Gleyzes:2014qga}, and GLPV theories themselves are closed under the same class of disformal transformations.
Further introduction of $X$-dependence into $A$ results in quadratic/cubic DHOST theories~\cite{BenAchour:2016fzp}.\footnote{To the best of our knowledge, it remains an open question whether these DHOST theories are closed under generic disformal transformations.}\footnote{Apart from this line of research, the authors of Ref.~\cite{Ezquiaga:2017ner} specified all the theories obtained via invertible disformal transformations from the Horndeski class in the language of differential forms.}
These papers explicitly showed that an invertible transformation does not change the number of physical DOFs.
In what follows, we show this for an arbitrary scalar-tensor theory as an application of our main theorem.
For the disformal transformation~\eqref{disformal2}, the linearization yields
	\be
	\bem \delta \ti{g}_{\mu\nu}\\ \delta \ti{\phi}\eem=\hat{P}\bem\delta g_{\al\beta}\\ \delta \phi\eem,~~~
	\hat{P}=\bem a^{\al\beta}_{\mu\nu}&\hat{b}_{\mu\nu}\\ 0&1\eem,
	\ee
where
	\begin{align}
	a^{\al\beta}_{\mu\nu}&\equiv \fr{1}{2}\bra{A_Xg_{\mu\nu}+B_X\na_\mu\phi\na_\nu\phi} \na^\al\phi\na^\beta\phi+A\delta^\al_{(\mu}\delta^\beta_{\nu)}, \\
	\hat{b}_{\mu\nu}&\equiv \bra{A_\phi g_{\mu\nu}+B_\phi\na_\mu\phi\na_\nu\phi}+\brb{2B\delta^\si_{(\mu}\na_{\nu)}\phi-\bra{A_Xg_{\mu\nu}+B_X\na_\mu\phi\na_\nu\phi}\na^\si\phi}\na_\si.
	\end{align}
The symmetrization for two indices is defined by $T^{\al\beta\ga\cdots}_{(\mu\nu)\si\cdots}\equiv \fr{1}{2}(T^{\al\beta\ga\cdots}_{\mu\nu\si\cdots}+T^{\al\beta\ga\cdots}_{\nu\mu\si\cdots})$, with $T^{\al\beta\ga\cdots}_{\mu\nu\si\cdots}$ being an arbitrary tensor.
Similarly, the inverse disformal transformation~\eqref{invdisformal} is linearized in the form of
	\be
	\bem \delta g_{\mu\nu}\\ \delta \phi\eem=\hat{Q}\bem\delta \ti{g}_{\al\beta}\\ \delta \ti{\phi}\eem,~~~
	\hat{Q}=\bem c^{\al\beta}_{\mu\nu}&\hat{d}_{\mu\nu}\\ 0&1\eem.
	\ee
Here, the matrix elements~$c^{\al\beta}_{\mu\nu}$ and $\hat{d}_{\mu\nu}$ can be written in terms of $(g_{\mu\nu},\phi)$ as
	\begin{align}
	c^{\al\beta}_{\mu\nu}&\equiv -\fr{1}{2A(A-XA_X+2X^2B_X)}\bra{A_Xg_{\mu\nu}+B_X\na_\mu\phi\na_\nu\phi} \na^\al\phi\na^\beta\phi+\fr{1}{A}\delta^\al_{(\mu}\delta^\beta_{\nu)}, \\
	\hat{d}_{\mu\nu}&\equiv -c^{\al\beta}_{\mu\nu}\hat{b}_{\al\beta}.
	\end{align}
As it should be, this $\hat{Q}$ defines the inverse matrix of $\hat{P}$:
One can check that
	\be
	\bem a^{\rho\si}_{\mu\nu}&\hat{b}_{\mu\nu}\\ 0&1\eem
	\bem c^{\al\beta}_{\rho\si}&\hat{d}_{\rho\si}\\ 0&1\eem
	=\bem c^{\rho\si}_{\mu\nu}&\hat{d}_{\mu\nu}\\ 0&1\eem
	\bem a^{\al\beta}_{\rho\si}&\hat{b}_{\rho\si}\\ 0&1\eem
	=\bem \delta^\al_{(\mu}\delta^\beta_{\nu)}&0\\ 0&1\eem.
	\ee
Now we confirm the equivalence between the old- and new-frame EOMs using the relation~\eqref{eomrel}.
Starting from a generic action~$S[\ti{g}_{\mu\nu},\ti{\phi}]$, the EOMs for the metric and the scalar field are derived as
	\be
	\ti{\mE}^{\mu\nu}\equiv\fr{\delta S}{\delta \ti{g}_{\mu\nu}}=0,~~~\ti{\mE}_\phi\equiv\fr{\delta S}{\delta \ti{\phi}}=0.\label{disfoldeom}
	\ee
On the other hand, if the action is written in terms of the new variables as $S'[g_{\mu\nu},\phi]$, the resulting EOMs are
	\be
	\mE^{\mu\nu}\equiv\fr{\delta S'}{\delta g_{\mu\nu}}=0,~~~\mE_\phi\equiv\fr{\delta S'}{\delta \phi}=0. \label{disfneweom}
	\ee
Then the relation~\eqref{eomrel} reads
	\be
	\bem \mE^{\al\beta}\\ \mE_\phi\eem=\hat{P}^\da\bem \ti{\mE}^{\mu\nu}\\ \ti{\mE}_\phi\eem,~~~
	\hat{P}^\da=\bem a^{\al\beta}_{\mu\nu}&0\\ \hat{b}_{\mu\nu}^\da&1\eem, \label{disfeomrel1}
	\ee
where the scalar equation~$\mE_\phi$ acquires higher derivative terms due to the contribution~$\hat{b}_{\mu\nu}^\da\ti{\mE}^{\mu\nu}$.
However, Eq.~\eqref{disfeomrel1} can be solved for the old-frame EOMs as
	\be
	\bem \ti{\mE}^{\al\beta}\\ \ti{\mE}_\phi\eem=\hat{Q}^\da\bem \mE^{\mu\nu}\\ \mE_\phi\eem,~~~
	\hat{Q}^\da=\bem c^{\al\beta}_{\mu\nu}&0\\ \hat{d}_{\mu\nu}^\da&1\eem, \label{disfeomrel2}
	\ee
which means that the lower-order EOMs~\eqref{disfoldeom} in the old frame can be recovered from the higher-order EOMs~\eqref{disfneweom} in the new frame.
The authors of Ref.~\cite{Arroja:2015wpa} gave the same result based on a heuristic approach, but our method has an advantage that the  equivalence between the old- and new-frame EOMs can be verified in a systematic manner.

%%%%%%%%%%%%%%%%%%%%%%%%%%%%%%%%%%%%%%%%%%
%%%%%%%%%%%%%%%%%%%%%%%%%%%%%%%%%%%%%%%%%%
\subsection{Mixing with derivatives of the metric}\label{ssec:scalar}

Contrary to the case of disformal transformations where only the metric is nontrivially transformed, here we consider a nontrivial transformation of the scalar field, namely,
	\be
	\ti{g}_{\mu\nu}=g_{\mu\nu},~~~\ti{\phi}=F(\phi;g_{\mu\nu},\pa_\la g_{\mu\nu},\pa_\la\pa_\si g_{\mu\nu},\cdots), \label{scalartrnsf}
	\ee
where $F$ is an arbitrary scalar quantity constructed without derivatives of $\phi$.
Note that the transformation~\eqref{scalartrnsf} generalizes the transformation~\eqref{f(R)} in \S \ref{ssec:ex2}.
If $\pa F/\pa\phi\ne0$, one can solve $F(\phi)=\ti{\phi}$ for $\phi$ in the form of
	\be
	\phi=\ti{F}(\ti{\phi};g_{\mu\nu},\pa_\la g_{\mu\nu},\pa_\la\pa_\si g_{\mu\nu},\cdots),
	\ee
which defines the inverse transformation as
	\be
	g_{\mu\nu}=\ti{g}_{\mu\nu},~~~\phi=\ti{F}(\ti{\phi};\ti{g}_{\mu\nu},\pa_\la \ti{g}_{\mu\nu},\pa_\la\pa_\si \ti{g}_{\mu\nu},\cdots). \label{invscalartrnsf}
	\ee

As we did in the previous section, we check the recoverability of the original-frame EOMs.
Following the prescription, Eq.~\eqref{scalartrnsf} is linearized as
	\be
	\bem \delta \ti{g}_{\mu\nu}\\ \delta \ti{\phi}\eem=\hat{P}\bem\delta g_{\al\beta}\\ \delta \phi\eem,~~~
	\hat{P}=\bem \delta^\al_{(\mu}\delta^\beta_{\nu)}&0\\ \hat{p}^{\al\beta}&F_\phi \eem,~~~\hat{p}^{\al\beta}\equiv \sum_s\fr{\pa F}{\pa(\pa_{(s)}g_{\al\beta})}\pa_{(s)},
	\ee
and its inverse transformation~\eqref{invscalartrnsf} as
	\be
	\bem \delta g_{\mu\nu}\\ \delta \phi\eem=\hat{Q}\bem\delta \ti{g}_{\al\beta}\\ \delta \ti{\phi}\eem,~~~
	\hat{Q}=\bem \delta^\al_{(\mu}\delta^\beta_{\nu)}&0\\ -\fr{1}{F_\phi}\hat{p}^{\al\beta}&\fr{1}{F_\phi} \eem=\hat{P}^{-1}.
	\ee
Now we find the relation between the old- and new-frame EOMs in the same manner as in the previous section:
	\be
	\bem \mE^{\al\beta}\\ \mE_\phi\eem=\hat{P}^\da\bem \ti{\mE}^{\mu\nu}\\ \ti{\mE}_\phi\eem,~~~
	\hat{P}^\da=\bem \delta^\al_{(\mu}\delta^\beta_{\nu)}&\hat{p}^{\da\al\beta} \\ 0&F_\phi \eem.
	\ee
In this case, the metric equation~$\mE^{\al\beta}$ becomes of higher order due to the contribution~$\hat{p}^{\da\al\beta}\ti{\mE}_\phi$.
Nevertheless, the new system has the same DOFs as the original one because the EOMs in the old frame can be recovered as
	\be
	\bem \ti{\mE}^{\al\beta}\\ \ti{\mE}_\phi\eem=\hat{Q}^\da\bem \mE^{\mu\nu}\\ \mE_\phi\eem,~~~
	\hat{Q}^\da=\bem \delta^\al_{(\mu}\delta^\beta_{\nu)}&-\hat{p}^{\da\al\beta}\fr{1}{F_\phi}\\ 0&\fr{1}{F_\phi} \eem.
	\ee

%%%%%%%%%%%%%%%%%%%%%%%%%%%%%%%%%%%%%%%%%%%%%%%%%%%%%%%%%%%%%%%%%%%%%%%%%%%%%%%%%%%%
%%%%%%%%%%%%%%%%%%%%%%%%%%%%%%%%%%%%%%%%%%%%%%%%%%%%%%%%%%%%%%%%%%%%%%%%%%%%%%%%%%%%
%	Conclusions
%%%%%%%%%%%%%%%%%%%%%%%%%%%%%%%%%%%%%%%%%%%%%%%%%%%%%%%%%%%%%%%%%%%%%%%%%%%%%%%%%%%%
%%%%%%%%%%%%%%%%%%%%%%%%%%%%%%%%%%%%%%%%%%%%%%%%%%%%%%%%%%%%%%%%%%%%%%%%%%%%%%%%%%%%
\section{Conclusions}\label{sec:conclusion}

Despite the common belief that an invertible transformation should not affect the number of physical DOFs, its validity is not clear if the transformation depends on derivatives of fields.
This is because the EL equations in the new frame consist of derivatives of order higher than in the original frame.
To address this issue, we showed in \S \ref{sec:mth} that there is a one-to-one correspondence between solutions in the two frames, which implies any pair of theories related by an invertible transformation has a common number of physical DOFs.

We also presented two examples of invertible transformations in scalar-tensor theories of gravity:
One is the disformal transformation, and the other is the transformation that contains derivatives of the metric.
We discussed these two types of transformations separately in \S \ref{sec:ST}, but one can further consider their compositions.
Such a composition of transformations generically take highly nontrivial form in which the metric and the scalar field are mixed with each other.
For example, the following transformation
	\be
	\ti{g}_{\mu\nu}=e^{2\phi}g_{\mu\nu},~~~\ti{\phi}=\phi+e^{-2\phi}(R+12X-6\,\Box \phi)
	\ee
has its inverse transformation and it is given by
	\be
	g_{\mu\nu}=e^{-2(\ti{\phi}-\ti{R})}\ti{g}_{\mu\nu},~~~\phi=\ti{\phi}-\ti{R}.
	\ee
As such, the space of invertible transformations on scalar-tensor theories has quite a rich structure.

Based on the main theorem, a class of theories which is obtained by an invertible transformation of some known healthy theories of gravity, such as the Horndeski class, could also be a class of unknown healthy theories.
In general, the resulting theories are seemingly by far beyond any known class due to the diversity of invertible transformations.
Although such theories themselves are not essentially new, they could provide some hint for new types of couplings in Lagrangian, and may lead us to construction of new theories that are not related to any known class by invertible transformations.

%%%%%%%%%%%%%%%%%%%%%%%%%%%%%%%%%%%%%%%%%%%%%%%%%%%%%%%%%%%%%%%%%%%%%%%%%%%%%%%%%%%%
%%%%%%%%%%%%%%%%%%%%%%%%%%%%%%%%%%%%%%%%%%%%%%%%%%%%%%%%%%%%%%%%%%%%%%%%%%%%%%%%%%%%

\acknowledgements{
H.M. is supported in part by
MINECO Grant No.~SEV-2014-0398,
PROMETEO II/2014/050,
Spanish Grant FPA2014-57816-P of the MINECO, and
European Union's Horizon 2020 research and innovation programme under the Marie Sk\l odowska-Curie grant agreements No.~690575 and No.~674896.
T.S. is supported in part by 
Japan Society for the Promotion of Science (JSPS) Grant-in-Aid for Young Scientists (B) No.~15K17632, 
Ministry of Education, Culture, Sports, Science and Technology (MEXT) Grant-in-Aid for Scientific Research on Innovative Areas ``New Developments in Astrophysics Through Multi-Messenger Observations of Gravitational Wave Sources'' No.~15H00777, and ``Cosmic Acceleration'' No.~15H05888.
The work of T.K. was supported
by MEXT-Supported Program for the Strategic Research Foundation at Private Universities, 2014-2017,
and by the JSPS Grants-in-Aid for Scientific Research No.~16H01102 and No.~16K17707.
H.M. thanks the Research Center for the Early Universe, where part of this work was completed.
}

%%%%%%%%%%%%%%%%%%%%%%%%%%%%%%%%%%%%%%%%%%%%%%%%%%%%%%%%%%%%%%%%%%%%%%%%%%%%%%%%%%%%
%%%%%%%%%%%%%%%%%%%%%%%%%%%%%%%%%%%%%%%%%%%%%%%%%%%%%%%%%%%%%%%%%%%%%%%%%%%%%%%%%%%%

\appendix*
\section{Removing field derivatives from field transformation}

We showed in \S \ref{ssec:main} that one can recover the original EOMs~$\mE_i^{(\phi)}=0$ from the new EOMs~$\hat{P}^\da{}^i_j\mE_i^{(\phi)}=0$ if the field transformation is invertible.
Here, $\hat{P}^\da{}^i_j$ contains derivative operators arising from derivatives in the field transformation, which is the origin of the nontriviality when proving the equivalence between the original- and new-frame EOMs.
To circumvent this problem, one may want to reduce the derivative-dependent transformation to a transformation without field derivatives by introducing auxiliary fields and Lagrange multipliers.
Naively, such a transformation allows us to obtain the new-frame EOMs in a more concise form and facilitates the proof of the equivalence between $\mE_i^{(\phi)}=0$ and $\hat{P}^\da{}^i_j\mE_i^{(\phi)}=0$.
However, it is actually not the case for the following reasons.

Let us go back to the Lagrangian~\eqref{gft} and the derivative-dependent transformation~\eqref{invtr}.
To remove derivatives from Eq.~\eqref{invtr}, we introduce auxiliary fields~$\chi^i_{(s)}\equiv\chi^i_{\mu_1\cdots\mu_s}$ with Lagrange multipliers~$\la_i^{(s)}\equiv\la_i^{\mu_1\cdots\mu_s}$ and obtain the modified Lagrangian as
	\be
	\ti{L}\equiv L(\phi^i,\pa_\mu\phi^i,\cdots,\pa_{(n)}\phi^i)+\sum_{s=1}^\ell\la_i^{(s)}(\chi^i_{(s)}-\pa_{(1)}\chi^i_{(s-1)}). \label{modL}
	\ee
Here, $\ell$ denotes the highest order of derivative in the transformation~\eqref{invtr}, and $\chi^i_{(0)}$ is understood as $\phi^i$.
The EL equations are
	\begin{align}
	\ti{\mE}_i^{(\phi)}&\equiv \mE_i^{(\phi)}+\pa_\mu\la_i^\mu=0, \label{modLeomphi} \\
	\ti{\mE}_{\chi^i_{(s)}}&\equiv \la_i^{(s)}+\pa_{(1)}\la_i^{(s+1)}=0, \label{modLeomchi} \\
	\ti{\mE}_{\la_i^{(s)}}&\equiv \chi^i_{(s)}-\pa_{(1)}\chi^i_{(s-1)}=0, \label{modLeomla}
	\end{align}
where $\la_i^{(\ell+1)}\equiv 0$.
Equations~\eqref{modLeomchi}, \eqref{modLeomla} yield $\la_i^{(s)}=0$, $\chi^i_{(s)}=\pa_{(s)}\phi^i$, respectively, and thus we obtain $\mE_i^{(\phi)}=0$ from Eq.~\eqref{modLeomphi}.
Now we formally replace the derivatives contained in the field transformation~\eqref{invtr} by $\chi^i_{(s)}$, namely,
	\be
	\psi^i=g^i(\phi^j,\chi^j_\mu,\cdots,\chi^j_{(\ell)}). \label{modinvtr}
	\ee
Assuming that $\chi^i_{(s)}$ and $\la_i^{(s)}$ remain unchanged when transformed into the new frame, Eq.~\eqref{modinvtr} defines an invertible transformation between extended field sets~$(\phi^i,\chi^i_{(s)},\la_i^{(s)})$ and $(\psi^i,\chi^i_{(s)},\la_i^{(s)})$ without field derivatives.
This is because the determinant of $J^i_j\equiv \pa g^i/\pa \phi^j$ is nonvanishing due to the invertibility of the field transformation~\eqref{invtr}.\footnote{The condition~$\det J^i_j\ne 0$ is only a necessary and not a sufficient condition for the field transformation to be invertible. For details, see Ref.~\cite{Motohashi:2016prk}.}
One may thus expect that (i) the relation between the old- and new-frame EOMs becomes clearer than considering the derivative-dependent transformation~\eqref{invtr}, and (ii) it would alleviate the proof of the equivalence between $\mE_i^{(\phi)}=0$ and $\hat{P}^\da{}^i_j\mE_i^{(\phi)}=0$.
Indeed, (i) is the case.
If we perform the transformation~\eqref{modinvtr} on the modified Lagrangian~\eqref{modL}, the variation of the action becomes as
	\begin{align}
	\delta\ti{S}&=\int d^Dx\delta\ti{L}=\int d^Dx\bra{\ti{\mE}_i^{(\phi)}\delta\phi^i+\ti{\mE}_{\chi^i_{(s)}}\delta\chi^i_{(s)}+\ti{\mE}_{\la_i^{(s)}}\delta\la_i^{(s)}} \nonumber \\
	&=\int d^Dx\brb{\ti{\mE}_i^{(\phi)}(J^{-1})^i_j\delta\psi^j+\bra{\ti{\mE}_{\chi^k_{(s)}}+\ti{\mE}_i^{(\phi)}(J^{-1})^i_j\fr{\pa g^j}{\pa \chi^k_{(s)}}}\delta\chi^k_{(s)}+\ti{\mE}_{\la_i^{(s)}}\delta\la_i^{(s)}}.
	\end{align}
Then, the resulting EL equations are
	\be
	(J^{-1})^i_j\ti{\mE}_i^{(\phi)}=0,~~~\ti{\mE}_{\chi^k_{(s)}}+(J^{-1})^i_j\fr{\pa g^j}{\pa \chi^k_{(s)}}\ti{\mE}_i^{(\phi)}=0,~~~\ti{\mE}_{\la_i^{(s)}}=0, \label{trmodLeom}
	\ee
which are obviously equivalent to Eqs.~\eqref{modLeomphi}-\eqref{modLeomla}, and thus the original EOMs~$\mE_i^{(\phi)}=0$.
However, (ii) is not the case since Eq.~\eqref{trmodLeom} does not address the equivalence between $\mE_i^{(\phi)}=0$ and $\hat{P}^\da{}^i_j\mE_i^{(\phi)}=0$.
Therefore, the idea of removing field derivatives from the field transformation does not lead to a simpler proof.

%%%%%%%%%%%%%%%%%%%%%%%%%%%%%%%%%%%%%%%%%%%%%%%%%%%%%%%%%%%%%%%%%%%%%%%%%%%%%%%%%%%%
%%%%%%%%%%%%%%%%%%%%%%%%%%%%%%%%%%%%%%%%%%%%%%%%%%%%%%%%%%%%%%%%%%%%%%%%%%%%%%%%%%%%

\bibliography{inv}

\end{document}